\documentclass{emulateapj}

\usepackage{graphicx}
\usepackage{color}

\newcommand{\uu}{$U$}%
\newcommand{\bb}{$B$}%
\newcommand{\vv}{$V$}%
\newcommand{\etal}{et~al.}%
\newcommand{\SN}{$S/N$}%
\newcommand{\figref}[1]{Figure~\ref{#1}}%
\newcommand{\secref}[1]{Section~\ref{#1}}%
\newcommand{\eqref}[1]{equation~(\ref{#1})}%
\newcommand{\mum}{$\mu$m}%
\newcommand{\bzero}{$\beta_{V,0}$}%
\newcommand{\av}{$A_V$}%
\newcommand{\magarcsec}{mag~arcsec$^{-2}$}%
\newcommand{\bmir} {(\bb\,$-$\,3.6\,\mum)}%
\newcommand{\fuvu} {(FUV\,$-$\,\uu)}%
\newcommand{\ffuvu}{0.188$\cdot$\fuvu\ $-$ 0.118}%
\newcommand{\lt}   {\,\ensuremath{<}\,}%
\renewcommand{\le} {\,\ensuremath{\leq}\,}%
\renewcommand{\ge} {\,\ensuremath{\geq}\,}%
\newcommand{\band} {\ \mbox{\bfseries and} \ }%
\newcommand{\bor}  {\ \mbox{\bfseries or} \ }%

\begin{document}

\title{Lifting the Veil of Dust from NGC\,0959: The Importance of a 
       Pixel-Based 2D Extinction Correction}

\shorttitle{Importance of 2D Extinction Correction}

\author{K.\ Tamura\altaffilmark{1}, 
	R.\ A.\ Jansen\altaffilmark{2,1},
	P.\ B.\ Eskridge\altaffilmark{2,3}, 
	S.\ H.\ Cohen\altaffilmark{2}, and
	R.\ A.\ Windhorst\altaffilmark{2,1}
}

\altaffiltext{1}{Department of Physics, Arizona State University, 
  Tempe, AZ 85287-1504, USA}

\altaffiltext{2}{School of Earth and Space Exploration, Arizona State 
  University, Tempe, AZ 85287-1404, USA}

\altaffiltext{3}{Department of Physics and Astronomy, Minnesota State
  University, Mankato, MN 56001, USA}

\email{ktamura@asu.edu}
\shortauthors{Tamura et al.}

\accepted{for publication in Astronomical Journal, April 14, 2010}

\begin{abstract}

We present the results of a study of the late-type spiral galaxy NGC\,0959,
\emph{before} and \emph{after} application of the pixel-based dust 
extinction correction
described in Tamura \etal\ 2009 (Paper~I).  \emph{Galaxy Evolution Explorer}
(\emph{GALEX}) far-UV (FUV) and near-UV (NUV), ground-based Vatican Advanced
Technology Telescope (VATT) \emph{UBVR}, and \emph{Spitzer}/Infrared Array
Camera (IRAC) 3.6, 4.5, 5.8, and 8.0\,\mum\ images are studied through pixel
Color-Magnitude Diagrams (pCMDs) and pixel Color-Color Diagrams (pCCDs).  We
define groups of pixels based on their distribution in a pCCD of \bmir\
versus \fuvu\ colors \emph{after} extinction correction.  In the same 
pCCD, we trace their locations \emph{before} the extinction correction was 
applied.  
This shows that selecting pixel groups is not meaningful when 
using colors uncorrected for dust.  
We also trace the distribution of the pixel groups on a pixel coordinate map
of the galaxy.  We find that the pixel-based (two-dimensional) extinction
correction is crucial to reveal the spatial variations in the dominant
stellar population, averaged over each resolution element.  Different types
and mixtures of stellar populations, and galaxy structures such as a
\emph{previously unrecognized} bar, become readily discernible in the
extinction-corrected pCCD and as coherent spatial structures in the pixel
coordinate map. 

\end{abstract}

\keywords{dust, extinction --- galaxies: individual (NGC\,0959) --- 
galaxies: spiral --- galaxies: stellar content --- galaxies: structure}

\section{Introduction}

Studying the distribution of different stellar populations within galaxies
is crucial to understand the formation and evolutionary history of galaxies. 
One of the obstacles that prevents us from directly observing these stellar
populations is the attenuation and reddening by dust that is randomly
distributed among, or projected along the line-of-sight toward, these
stellar populations.  This has remained a major problem for nearly a century
\citep[e.g.,][]{trumpler30, mathis77, viallefond82, caplan85, witt92,
roussel05, driver08}.  While the main effect of dust is to attenuate the
light emitted by stars, dust also scatters light out of the line-of-sight,
and may re-direct light from nearby regions into our line-of-sight
\citep[e.g.,][]{witt92, witt96, witt00}.  The impact of extinction---the
wavelength-dependent net effect of absorption and scattering---correlates
with the spatial distribution of star-forming regions
\citep[e.g.,][]{waller92, deo06}, as it depends on both the geometry of the
dust distribution \citep[e.g.,][]{elmegreen80, walterbos88, calzetti94,
witt96, witt00} and on the physical and chemical properties of the dust
\citep[e.g.,][]{vanhouten61, witt92, whittet01, whittet04}.  On typical
observational scales, the dust will be intermixed with the stellar
populations as thin layers, filaments, and dense clumps.  Even though some
light may be scattered into the line-of-sight from the rear and from regions
near the line-of-sight, most of the extinction by far will be due to dust
that is distributed \emph{in front} of the stellar populations of interest
\citep{byun92}. 

Popular methods to measure dust extinction, such as the use of Hydrogen
recombination line ratios \citep[e.g.,][]{rudy84, scoville01, calzetti05,
calzetti07, kennicutt09}, the ultraviolet (UV) spectral slope
\citep[e.g.,][]{calzetti94, kong04}, or ratios of the UV and total infrared
fluxes \citep[e.g.,][]{buat96, calzetti00, boissier04, boissier05}, have
various constraints and limitations \citep[e.g.,][]{xu96, petersen97,
regan00, price02, boissier04, rieke04}.  In \citet[][hereafter
Paper~I]{tamura09a}, we introduced a new method to measure the dust 
extinction in galaxies, and demonstrated that the spatial distribution of 
dust extinction in a late-type spiral galaxy, NGC\,0959, can be mapped 
using only the \emph{flux ratio of optical \vv-band and mid-infrared 
(mid-IR) 3.6\,\mum\ images}.  In the present paper, we
present the results of a multi-wavelength (UV--optical--mid-IR) study of the
effect of the pixel-based two-dimensional extinction correction of Paper~I
on color composite images, a pixel Color--Color Diagram (pCCD), and a pixel
coordinate map of NGC\,0959.  These results show that a detailed pixel-based
extinction correction is essential to reveal the nature and distribution of
stellar populations in galaxies.

\section{Assumptions about Dust Extinction}

Even though the effects of dust attenuation dominate over those by
scattering, scattering cannot be ignored completely.  \citet{witt96, witt00}
show that, depending on dust geometry, the effect of scattering can be
significant.  Among the modeled dust geometries, their CLOUDY model
corresponds closest to the dust geometry we assumed in Paper~I: a geometry
where thin layers, filaments, and dense clumps of dust are intermixed with
the stars.  \citet[][their Figure~8]{witt00} find that this model shows the
smallest effect from scattering, with $F_{\rm scat}$/$F_*$\,$\lesssim$\,0.1
\citep[or $\Delta\mu\lesssim$\,0.1\,\magarcsec, in agreement with][]{byun92}
for all modeled wavelengths ($\lambda$\,$>$\,0.1\,\micron) and
metallicities, even for very clumpy dust distributions. 

While \citet{witt96, witt00} model the effect of dust attenuation and
scattering, \citet{gordon03}---and references therein---measured the
UV--near-IR extinction toward different regions within the Milky Way (MW),
the Large Magellanic Cloud (LMC), and the Small Magellanic Cloud (SMC). 
Since the effects of attenuation and scattering cannot be separated, the
observed extinction curves in \citet{gordon03} are the total effect of dust. 
Similarly, the extinction measured in Paper~I represents the total, or net,
effect of both attenuation and scattering. 

One needs to adopt an extinction curve to scale the measured visual 
extinction, $A_V$, to the extinction $A_{\lambda}$ in other bandpasses. 
There appears to be a systematic change from an SMC \citep{gordon98} 
via an LMC \citep{misselt99} to a MW type \citep{clayton00, valencic03} 
extinction curve, which---while controversial and 
unproven---is often taken to reflect the difference in average metallicity
and/or morphology of these galaxies.  Even within a single galaxy, however,
there is significant variation between extinction curves derived for 
different sightlines toward individual stars or regions
\citep[e.g.,][]{gordon98, misselt99, clayton00, valencic03, gordon03}.  
The rapid decrease
in attainable spatial resolution with increasing distance renders similar
measurements impossible for galaxies beyond the Local Group.  Based on
its overall morphology and luminosity, we assumed in Paper~I that an
LMC-like extinction curve---actually the ``LMC2 supershell'' extinction
curve of \citet{gordon03}---would be more appropriate for NGC\,0959 than
either an SMC or MW-type extinction curve.  At the $\simeq$\,10~Mpc 
distance of NGC\,0959,
we expect that the mixing of light from neighboring stellar populations on
$\sim$250\,pc scales will tend toward a mean similar to that for the LMC2
supershell, even when the extinction toward individual stellar populations
might be better characterized by an SMC or MW-type extinction curve.
Adopting the SMC bar \citep{gordon03} or an average MW-extinction curve,
would over- or under-estimate the extinction at shorter wavelengths, 
respectively.

\section{Data-Set and Preparation}

The galaxy selected for this pilot study, NGC\,0959, is a late-type spiral
galaxy \citep[Sdm;][]{devaucouleurs91} at a distance of
$D$\,=\,9.9\,$\pm$\,0.7\,Mpc \citep{mould00} with an inclination of
$\sim$\,50$\degr$ \citep{esipov91}.  This galaxy has been observed with
\emph{GALEX} \citep{martin05, morrissey07} in the FUV and near-UV (NUV),
with the Vatican Advanced Technology Telescope (VATT) in \emph{UBVR} from
the ground \citep[][]{taylor05}, and with \emph{Spitzer}/IRAC in the 3.6,
4.5, 5.8, and 8.0\,\mum\ filters \citep{fazio04}.  Near-IR $JHK_s$ images
are also available from the Two Micron All Sky Survey \citep[2MASS;][]{
skrutskie06}, but these near-IR images are too shallow for our purpose, and
therefore excluded from the analysis.  All displayed images and pixel-maps
are rotated such that North is up and East is to the left.  The surface
brightness and colors are in units of \magarcsec\ and mag, respectively, and
are on the AB magnitude system \citep{oke74, oke83}. 

To perform the pixel-based analysis, all images are convolved and re-sampled
to a matching point spread function (PSF) and pixel scale.  Among all images
from the different telescopes and instruments of interest, the coarsest
pixel scale and PSF are 1\farcs5 pixel$^{-1}$ and $\sim$\,5\farcs3 FWHM,
respectively, for the \emph{GALEX} NUV image.  
IDL\footnote[4]{
  IDL is distributed by ITT Visual Information Solutions (Research System
  Inc.), Boulder, Colorado: \texttt{http://rsinc.com/idl/}
} function ``\texttt{frebin}'' (with its flux conserving option switched
on) and IRAF\footnote[5]{
  IRAF is distributed by National Optical Astronomy Observatory (NOAO),
  which is operated by the Association of Universities for Research in 
  Astronomy (AURA), Inc., under cooperative agreement with National Science
  Foundation (NSF): \ \texttt{http://iraf.net/}
} 
routine ``\texttt{gauss}'' (with a round two-dimensional Gaussian
convolution kernel) are used to match the pixel scale and PSF at all
wavelengths, while conserving the total amount of surface flux.  We then
select only those pixels that have a signal-to-noise (\SN) $\geq$\,3.0 in
all filters for further analysis, ensuring reliable pixel surface brightness
measurements and colors.  Below, we briefly describe the data used in this
analysis.  For the details of the data preparation and the method of
extinction correction we refer the reader to Paper~I.

\subsection{\emph{GALEX} Images} 

The \emph{GALEX} FUV and NUV images are obtained from the Multi-Mission 
Archive at the Space Telescope Science Institute\footnote[6]{
  Galaxy Evolution Explorer, GR4 Data Release (May 1, 2009): \ 
  \texttt{http://galex.stsci.edu/GR4/}
} 
(MAST).  NGC\,0959 was observed in the \emph{GALEX} All-sky Imaging Survey
\citep[AIS;][]{martin05} and the \emph{GALEX} Nearby Galaxy Survey
\citep[NGS;][]{gildepaz07}.  Since the galaxy is barely visible in the AIS
images, we will use only the deeper ($\sim$\,1695 sec) NGS data for our
analysis, focusing only on a 7\farcm5\,$\times$\,7\farcm5 image section
centered on NGC\,0959.  Since the average FWHM of stars near NGC\,0959 is
larger in the NUV image \citep[e.g.,][]{martin05}, we convolve the FUV image
to match the PSF in the NUV.

\subsection{Ground-based Images}

Ground-based \emph{UBVR} images were obtained by \citet{taylor05} with the 
direct CCD imager at the VATT at {Mt.\,Graham} International Observatory 
(MGIO) in Arizona.  These flux calibrated images are available through the
NASA/IPAC Extragalactic Database\footnote[7]{
  NASA/IPAC Extragalactic Database: \
  \texttt{http://nedwww.ipac.caltech.edu/} 
} 
(NED).  For a detailed description of this data-set, we refer the reader to
\citet{taylor05}, and references therein.  The effective exposure times in
\emph{U}, \emph{B}, \emph{V}, and \emph{R} are 1200, 600, 480, and 360\,sec,
respectively.  The native resolution and pixel scale are $\sim$1\farcs3
(FWHM) and 0\farcs37 pixel$^{-1}$.  These images are registered, convolved
and resampled to match the orientation, PSF, and pixel scale of the
\emph{GALEX} NUV image. 

\vspace{15mm}
\subsection{\emph{Spitzer}/IRAC Images}

The \emph{Spitzer}/IRAC 3.6--8.0\,\mum\ pipeline-product images were
obtained from the \emph{Spitzer} Archive\footnote[8]{
  \emph{Spitzer} Science Center (SCC) Data Archives/Analysis: \ 
  \texttt{http://ssc.spitzer.caltech.edu/archanaly}
} 
via \texttt{Leopard}.  For each filter, the mosaiced image was created with
an effective exposure time of 26.8\,sec.  The native pixel scale is 1\farcs2
pixel$^{-1}$, and the effective resolution (FWHM) ranges from $\sim$2\farcs2
at 3.6\,\mum\ to $\sim$2\farcs3 at 8.0\,\mum.  Like for the VATT images, we
match the orientations, PSF, and pixel scale of the IRAC images to those of
the \emph{GALEX} NUV image.

\section{Constructing an Extinction Map of NGC\,0959}

In Paper~I, we estimated the visual dust extinction (\av\ in \magarcsec)
measured over each 1\farcs5$\times$1\farcs5 $\simeq$ 72$\times$72\,pc$^2$
pixel in NGC\,0959 as follows.  From histograms of the observed visual to
3.6\,\mum\ flux ratio ($f_V$/$f_{\hbox{\scriptsize 3.6\mum}}$) in each
pixel, we estimate the intrinsic extinction-free flux ratios (\bzero) for
two groups of pixels: pixels apparently dominated by the light from younger
and pixels apparently dominated by the light from older stellar populations. 
These are separable based on their distribution pattern in a pCMD of the
observed $\mu_V$ versus ($U$\,$-$\,3.6\,\mum) color.  
Since the mid-IR 3.6\,\mum\ flux is
assumed to be minimally affected by the dust, and hence usually treated as
extinction-free \citep[e.g.,][]{fazio04, willner04}, \av\ in each pixel can
be inferred from the difference between the \emph{observed}
($f_{V,\rm obs}$) and \emph{estimated} extinction-free (i.e., 
$f_{V,\rm 0}$\,=\,$\beta_{V,\rm 0} \times f_{\rm 3.6\mu m,obs}$) \vv-band
fluxes.  For further details of the method, we refer the reader to Paper~I.
The extinctions in other bandpasses (\emph{GALEX} NUV and FUV, and optical
\emph{U}, \emph{B}, and \emph{R} bands) are then scaled from \av\ using the
adopted LMC2 supershell extinction curve of \citet{gordon03}. 

\begin{figure} 
  \centerline{
  \includegraphics[width=0.49\textwidth,angle=0,clip=]{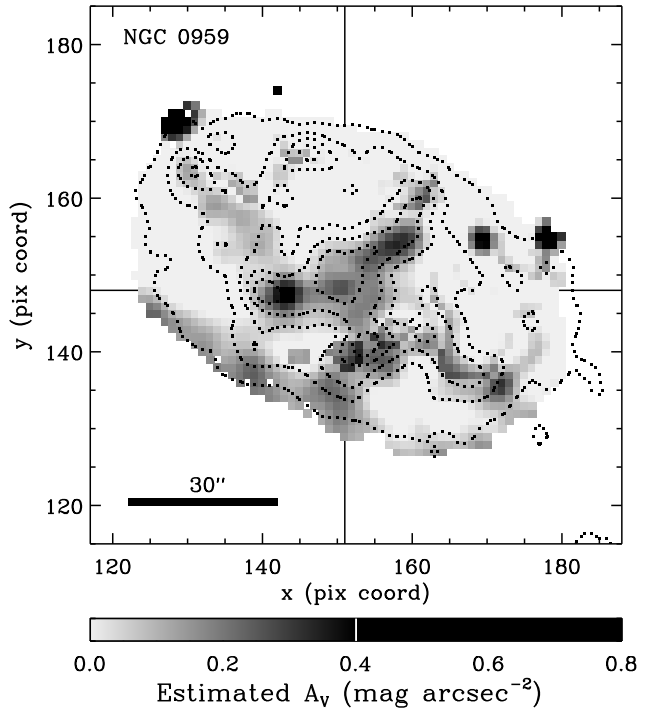}
  }
  \caption{\footnotesize
Spatial distribution of the pixel-averaged visual dust extinction, \av.
Extinction values map onto grayscales from light-gray
(\av\,=\,0) to black (\av\,=\,0.4\,\magarcsec; indicated by a white vertical
line in the color bar).  The few pixels with 0.4\,$\le$\,\av\,$\le$\,0.8
\magarcsec\ are also rendered as black.  Dotted contours trace the
\emph{Spitzer}/IRAC 8.0\,\mum\ PAH emission and show good agreement with the
visual extinction estimates, which were inferred from the observed \vv-band
and 3.6\,\mum\ fluxes only.
The map measures $\sim$135\arcsec$\times$120\arcsec, has a plate
scale of 1$\farcs$5\,pixel$^{-1}$, and has North up and East to the left.
  }\label{avmap}
\end{figure}

In \figref{avmap}, the extinction map of NGC\,0959 thus produced, darker
grayscales correspond to higher values of \av.  The grayscales saturate for
\av\,$\geq$\,0.4\,\magarcsec\ (indicated by the white vertical line in the
color bar) to enhance the visibility of the lower \av\ values in the galaxy,
and so differ from that of Figure~10(\emph{a}) in Paper~I.  The maximum
extinction measured in this galaxy, averaged over a pixel, is $A_{V,\rm
max}$\,$\simeq$\,0.8\,\magarcsec. 

\section{Lifting the Veil of Dust from NGC\,0959}

\subsection{Color Composite Images}\label{color_comp}

\begin{figure*} 
  \centerline{
  \includegraphics[height=0.49\textwidth,angle=0,clip=]{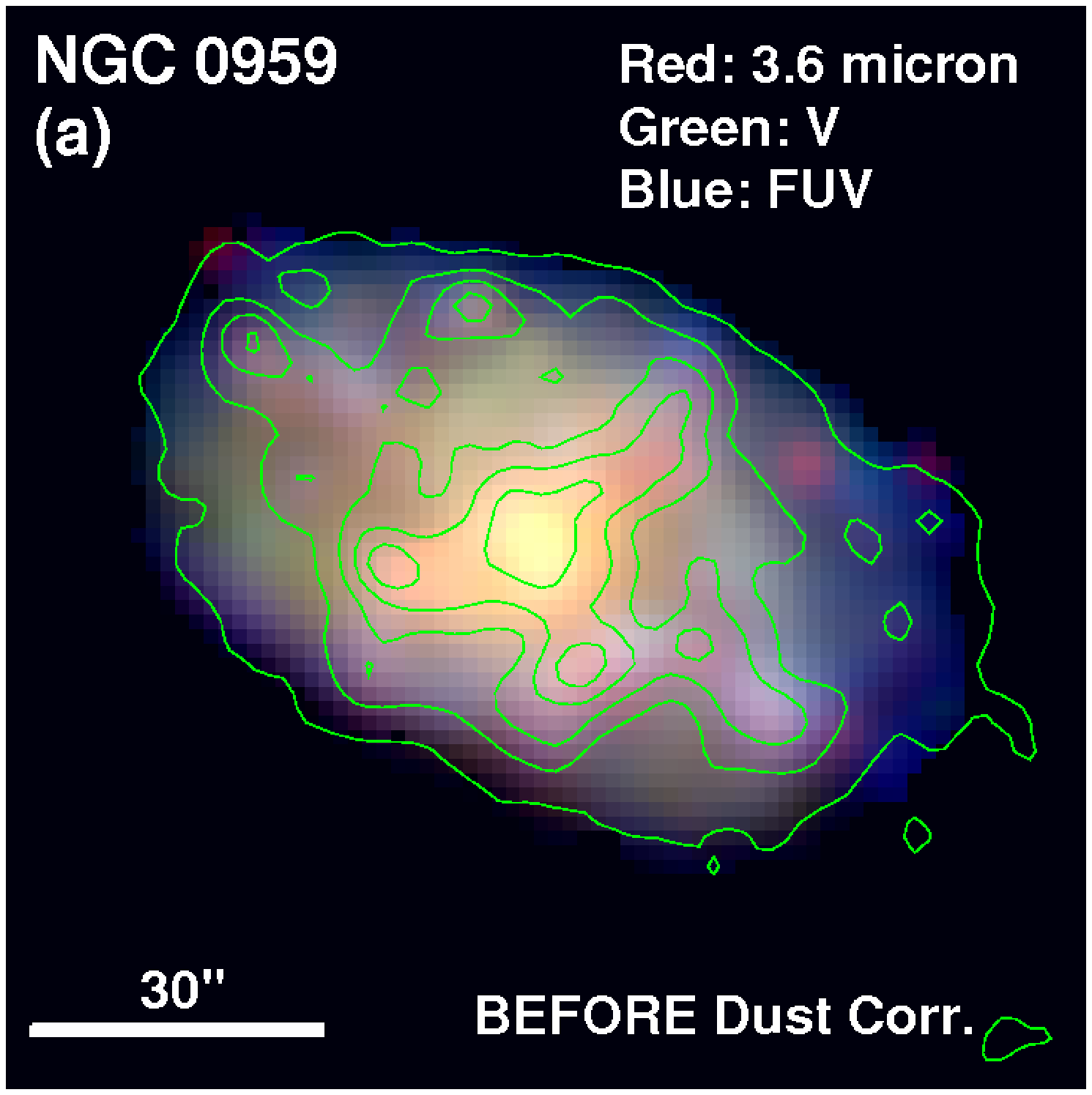}
  \hspace{3mm}
  \includegraphics[height=0.49\textwidth,angle=0,clip=]{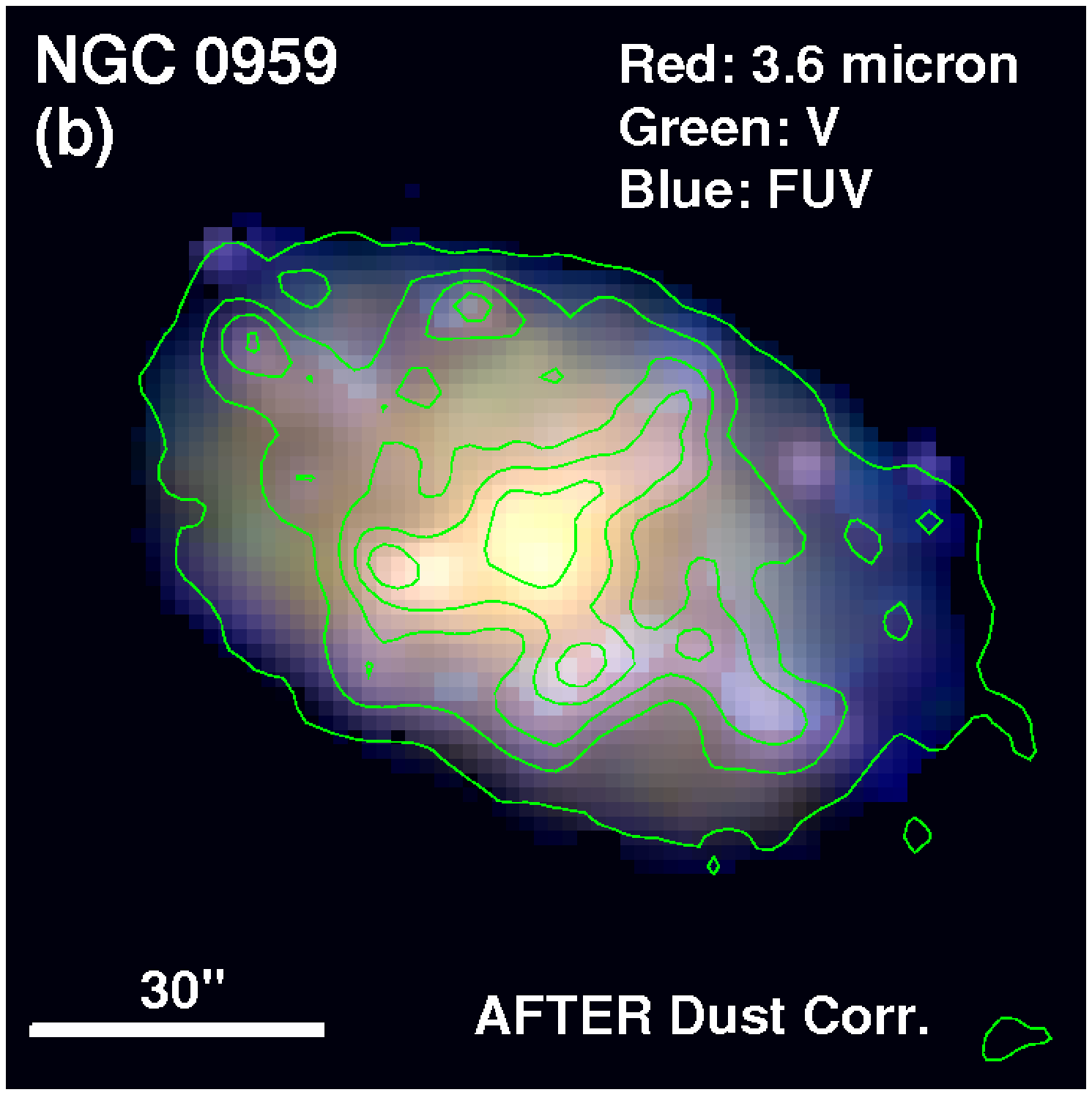}
  }
  \caption{\footnotesize
Color composite images of NGC\,0959 using images from \emph{Spitzer}/IRAC
3.6\,\mum\ (red), VATT \vv\ (green), and \emph{GALEX} FUV (blue) at
\emph{GALEX} resolution (\emph{a}) \emph{before} and (\emph{b}) \emph{after}
application of our pixel-based dust-extinction correction (Paper~I).  The
\emph{Spitzer}/IRAC 8.0\,\mum\ emission is over-plotted as green contours.
Both color composites are created using the same color stretch.  Regions
corresponding to high \av\ values in \figref{avmap} clearly become much
bluer after extinction correction (panel \emph{b}).  This is especially
clear in the blue star-forming knots which generally coincide with peaks in
the \emph{Spitzer} 8.0\,\mum\ emission.
  }\label{colimg}
\end{figure*}

To visually (qualitatively) investigate the effect of our pixel-based
extinction correction on an image of NGC\,0959, we first construct two color
composites of the galaxy, composed of the \emph{Spitzer}/IRAC 3.6\,\mum\
(red channel), the ground-based \vv\ (green channel), and the \emph{GALEX}
FUV (blue channel) images.  Figures~\ref{colimg}(\emph{a}) and
\ref{colimg}(\emph{b}) show the color composites \emph{before} and
\emph{after} extinction correction.  The image resolutions are matched to
that of the \emph{GALEX} NUV image.  For easy comparison, both images were
created using the \emph{same} color stretch, and with the IRAC 8.0\,\mum\
contours over-plotted. 

In \figref{colimg}(\emph{a}), there are several regions in the galaxy that
appear much redder than other parts of the galaxy.  These regions include:
(a) the southern half of the galaxy, especially along the southern ``edge''
of the galaxy disk (as defined by our $S/N_{\rm min}$\,=\,3.0 requirement in
each of the filters); (b) the strong 8.0\,\mum\ emission region that is
running from NW to SE of the galaxy through its center, appearing especially
redder at its northern end point; and (c) some localized regions at the
north-western and eastern edge of the galaxy disk.  Since dust features are
not resolved at \emph{GALEX} resolution, we cannot tell whether these pixels
in \figref{colimg}(\emph{a}) are red due to extinction, or are dominated by
intrinsically red stellar populations.  From the distribution of the
estimated visual dust extinction \av\ (\figref{avmap}),
however, we infer that these red pixels are indeed red due to intervening 
dust at these locations.

\begin{figure} 
  \centerline{
  \includegraphics[width=0.49\textwidth,angle=0,clip=]{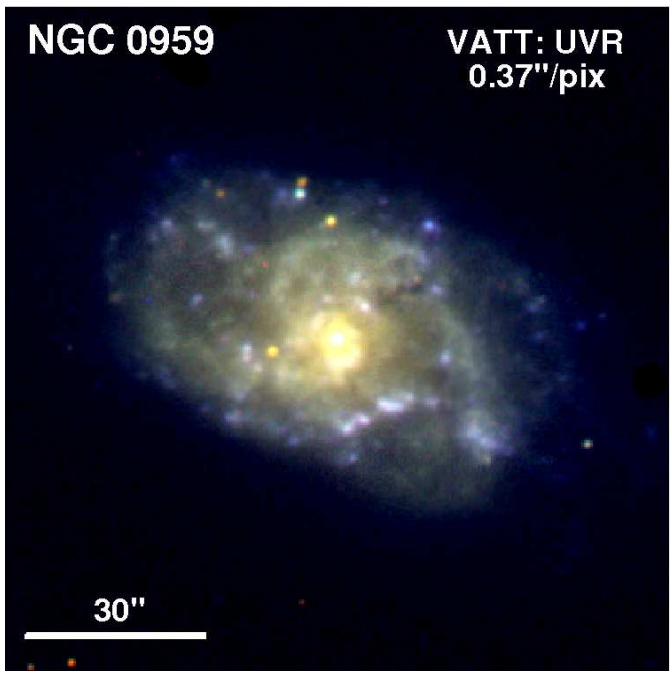}
  }
  \caption{\footnotesize
Color composite image of NGC\,0959 using VATT \emph{U}, \emph{V}, and \emph{R}
images at their native ground-based resolution (FWHM\,$\simeq$\,1\farcs3)
and pixel scale (0\farcs37\,pixel$^{-1}$) and \emph{without} any extinction
correction.  Some dust features and blue star forming regions are readily
discernible.
  }\label{vattimg}
\end{figure}

After applying our extinction correction, \figref{colimg}(\emph{b}) shows 
that many of the strikingly red pixels in \figref{colimg}(\emph{a}) indeed 
have become bluer.  
Especially regions (b) and (c) have become much bluer than
before, while some of the regions (a) still seem to have relatively red
colors compared to other parts of NGC\,0959's disk.  There are some other
regions which have become much bluer as well---mostly located in the SW
region and northern regions of the galaxy.  These regions already had a
relatively blue hue in \figref{colimg}(\emph{a}), but become much bluer and
brighter after extinction correction.  They likely are actively star-forming
regions. 

\begin{figure*} 
  \centerline{
  \includegraphics[width=0.49\textwidth,angle=0,clip=]{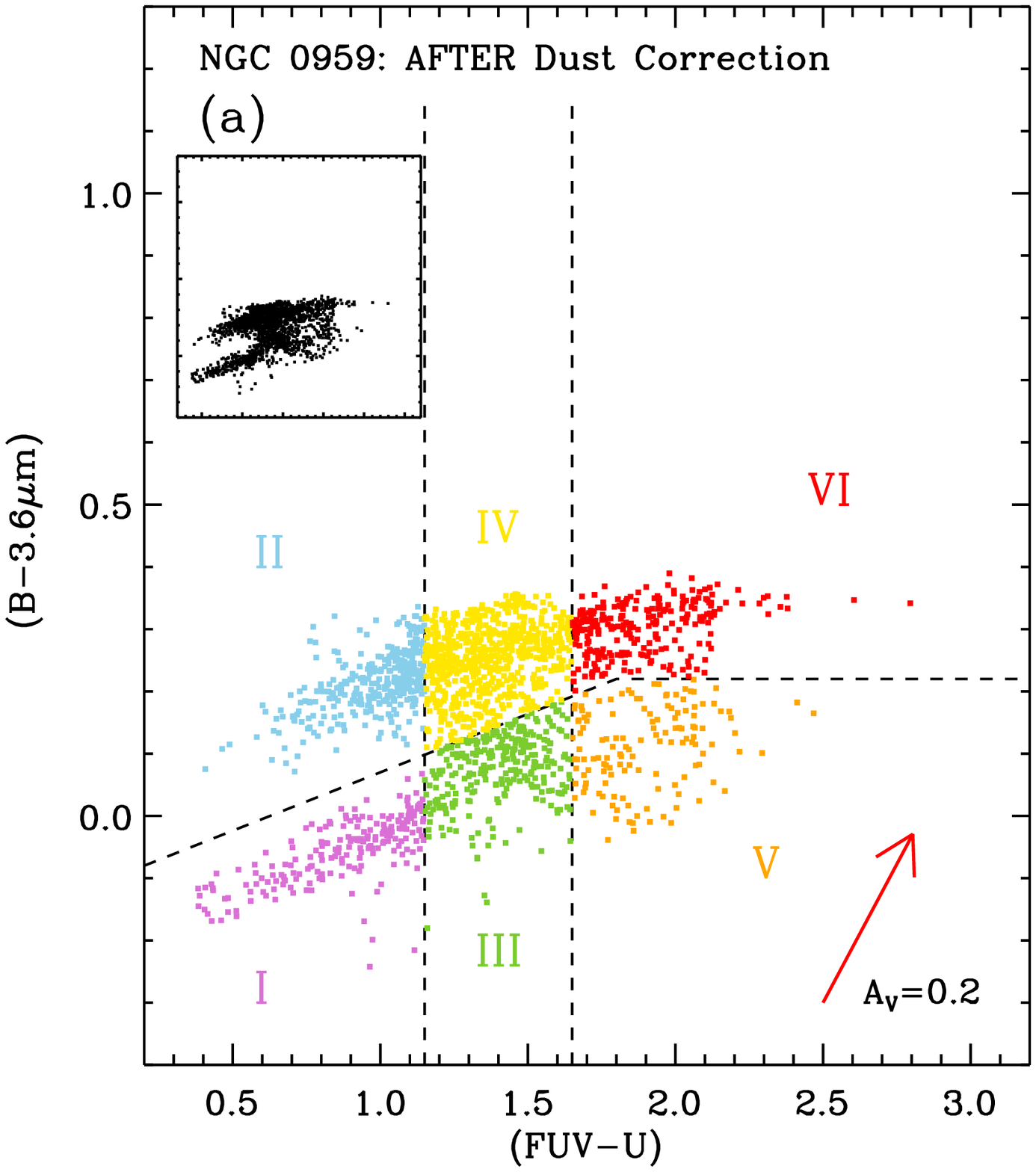} 
  \includegraphics[width=0.49\textwidth,angle=0,clip=]{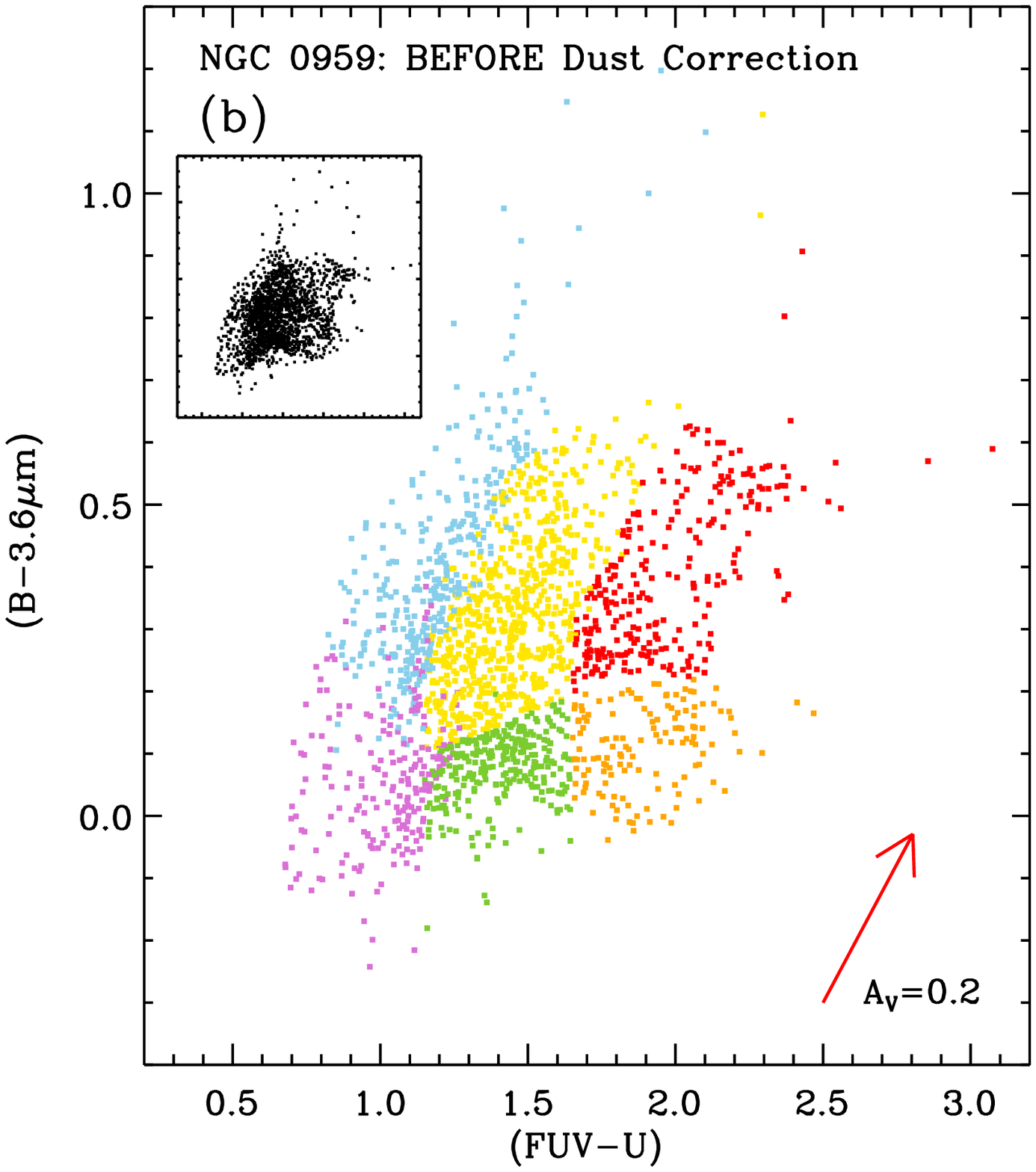}
  }
  \caption{\footnotesize
Pixel Color-Color Diagrams (pCCDs) of \bmir\ versus \fuvu\ color for
NGC\,0959.
Panel (\emph{a}) shows the pCCD \emph{after} our extinction correction (see
Paper~I) has been applied, while panel (\emph{b}) shows the pCCD using the
\emph{observed} pixel colors.  The 25$^{\rm th}$ and 75$^{\rm th}$
percentile uncertainties in color are $\sim$\,0.05 and $\sim$\,0.08 mag in
\bmir, and $\sim$\,0.01 and $\sim$\,0.02 mag in \fuvu, respectively.
Reddening vectors corresponding to \av\,=\,0.2 \magarcsec\ are drawn in the
bottom right corners.  Pixel groups were selected in panel (\emph{a}) and
retain their group assignment and color-coding in panel (\emph{b}).  Small
insets show the pCCDs, plotted over the same
color ranges as the main panels, but omitting the color coding that might
guide the reader's eye.  While the pCCD before extinction correction shows
no prominent well-separated features in color-color space, distinct
sequences and pixel groupings appear \emph{after} application of our
extinction correction (panel \emph{a}).
  }\label{pccd}
\end{figure*}

To see whether these red and blue regions in \figref{colimg} actually
correspond to dust features and SF-regions, we created a third color
composite image with a higher spatial resolution from the ground-based
\emph{UVR} images at their native resolution and pixel scale of
FWHM\,$\simeq$\,1\farcs3 (matched across \emph{UVR}) and 0\farcs37
pixel$^{-1}$.  \figref{vattimg} shows that the blue regions that become
bluer and brighter are indeed likely SF-regions.  It also shows that the
redder pixels in region (b) described above seem to be caused by thick dust
lanes running from the NW to SE in the galaxy.  It is hard to see whether
regions (a) and (c) are caused by a dust lane or not, but regions (a) seems
to be distributed around bluer clumps in \figref{vattimg}.  The localized
regions (c) do not appear to be unrelated background or foreground objects. 
Such objects usually have colors sufficiently different from those of the
genuine galaxy pixels, that they would appear as a distinct and separate
branch or grouping of pixels in pCMDs and pCCDs.  Since we found no such
feature in the examined pCMDs and pCCDs for NGC\,0959 (e.g., \figref{pccd}), 
we treat these regions as parts of the galaxy.

\subsection{A ($B$\,$-$\,3.6\,\mum) versus (FUV\,$-$\,$U$) pCCD}

To study the effect of our pixel-based extinction correction in a more
quantitative way, we examined pCMDs and pCCDs using various combinations of
images from FUV through 8.0\,\mum\ for significant features or groupings in
the pixel-distribution.  A distinct grouping of pixels in these diagrams
should indicate that those pixels are dominated by the same or a similar 
mix of stellar populations.  Among different diagrams examined 
(not shown here), a pCCD of \bmir\ versus \fuvu\ color \emph{after} 
extinction correction (\figref{pccd}(\emph{a})) was selected, because it 
clearly shows distinct tracks and groupings of pixels.  In particular, 
distinct red and blue sequences can be identified. 

The uncertainties in the colors at the 25$^{\rm th}$ and 75$^{\rm th}$
percentiles are $\sim$\,0.01 and $\sim$\,0.02\,mag in \fuvu, and
$\sim$\,0.05 and $\sim$\,0.08\,mag in \bmir, respectively.  Since the gap
of $\sim$\,0.2\,mag at the bluer end of \fuvu\ color between the two
sequences in \figref{pccd}(\emph{a}) is much larger than the photometric
uncertainties in the colors, this separation of pixels is not a random
effect.  The sequences are also not the result of differences in residual
reddening, as can be seen by comparing the directions of the sequences with
that of the reddening vector drawn in \figref{pccd}(\emph{a}).  The red and
blue sequences are connected in the
1.15\,$\lesssim$\,\fuvu\,$\lesssim$\,1.65\,mag color range by pixels with
intermediate colors.  At colors redder than \fuvu\,$\simeq$\,1.65\,mag, the
pixels again somewhat separate into sequences with redder and bluer \bmir\
colors.

\begin{table*}[!t]    
\footnotesize
\caption{Group Boundaries in the \bmir\ versus \fuvu\ pCCD of
	\figref{pccd}(\emph{a})\label{tbl_1}\protect\rule[-10pt]{0pt}{10pt}
}
\begin{tabular}{llc}
\hline\hline\\[-10pt]
Pixel Group & Color Code & Group Boundary Definition (mag)\\[2pt]
\hline\\[-8pt]
Group I   & Purple & \fuvu\ \lt\ 1.15             \\
          &        & \bmir\ \lt\ \ffuvu                        \\[6pt]
Group II  & Blue   & \fuvu\ \lt\ 1.15             \\
          &        & \bmir\ \ge\ \ffuvu                        \\[6pt]
Group III & Green  &   1.15 \le\ \fuvu\ \lt\ 1.65 \\
          &        & \bmir\ \lt\ \ffuvu                        \\[6pt]
Group IV  & Yellow &   1.15 \le\ \fuvu\ \lt\ 1.65 \\
          &        & \bmir\ \ge\ \ffuvu                        \\[6pt]
Group V   & Orange & \fuvu\ \ge\ 1.65             \\
          &        & \bmir\ \lt\ 0.22 \band \bmir\ \lt\ \ffuvu \\[6pt]
Group VI  & Red    & \fuvu\ \ge\ 1.65             \\
          &        & \bmir\ \ge\ 0.22 \bor \bmir\ \ge\ \ffuvu  \\[4pt]
\hline\\[12pt]
\end{tabular}
\end{table*}

As can be seen in various SED models \citep[e.g.,][]{bruzual03, anders03,
maraston05, kotulla09}, the \fuvu\ color is very sensitive to age for the
youngest stellar populations.  Since most of the FUV flux is emitted by
young, massive OB-stars, combining a \emph{GALEX} and an optical broad-band
filter provides strong age constraints \citep[e.g.,][]{kaviraj07}.  The
\bmir\ color was selected empirically, while examining different
combinations of colors.  The IRAC 3.6\,\mum\ bandpass is commonly used as a
stellar mass distribution tracer \citep[e.g.,][]{ willner04}, because it is
associated more with the distribution of redder and older stars.  Since the
optical \bb-band is generally sensitive to younger stars, the \bmir\ color
can be used to distinguish mixtures of stellar populations with and without
significant recent high-mass star formation.  The combination of \fuvu\ and
\bmir\ colors in \figref{pccd}(\emph{a}), therefore, provides a powerful
diagnostic of the recent star formation history averaged over a pixel.

\vspace{15mm}
\subsection{Definition of Pixel Groups}

Based on the distinct red and blue sequences, as well as transition regions
identified in \figref{pccd}(\emph{a}), we separated pixels into six different
groups (color-coded in \figref{pccd}(\emph{a})).  The group boundaries
(selection criteria) are summarized in Table~1.  The exact location of the
boundaries between some of these pixel groups is somewhat arbitrary, but is
motivated by the following considerations. 

The Group~I pixels (color-coded purple) reside on the blue part of the blue
sequence, and have blue colors in both \bmir\ and \fuvu.  The dominant
stellar populations are very young and massive OB-stars, while older stellar
populations contribute little to the total flux in these pixels.  The \bmir\
and \fuvu\ colors become redder as these young stellar populations age and
the number of remaining OB-stars decreases. 

On the other extreme, the Group~VI pixels (color-coded red) have red colors
in both \bmir\ and \fuvu.  These pixels are therefore dominated by old,
quiescent, ``red-and-dead'' stellar populations and do not contain a
detectable fraction of young, massive stars. 

The Group~II pixels (color-coded blue) have the same blue \fuvu\ color range
as Group~I, but redder \bmir\ colors.  The blue \fuvu\ color implies the
presence of OB-stars.  The redder \bmir\ color indicates that Group~II
pixels have a non-negligible contribution to the total light from older
(underlying or superposed along the line-of-sight) stellar populations. 

Group~V pixels (color-coded orange) cover the same, red, \fuvu\ color range
as Group~VI, indicating that they too lack a detectable fraction of young,
massive OB-stars.  Yet, their bluer \bmir\ color suggests that the flux in
these pixels is dominated by light from intermediate age stellar
populations. 

The Group~III and Group~IV pixels (color-coded green and yellow,
respectively) are located between these extreme cases.  The light in these
pixels is likely dominated by stellar populations in transition between the
extreme groups along either blue or red sequences, or between the blue and
the red sequence, or represents a mixture of stellar populations (from
unresolved adjacent regions or regions superposed along the line-of-sight)
with different star formation histories.

\subsection{Effect of Pixel-Based Extinction Correction}

\figref{pccd}(\emph{b}) shows the pCCD of \bmir\ versus \fuvu\ color
\emph{before} the application of the extinction correction.  For each pixel,
the color coding was preserved from that in \figref{pccd}(\emph{a}).  This
allows us to track the effect of applying the extinction correction in this
color--color space.  Interestingly, as shown in the inset of
\figref{pccd}(\emph{b}), once the color coding is removed, obvious features
like the blue and red sequences of \figref{pccd}(\emph{a}) are no longer
discernible.  The Group~I and II pixels, for example, which were clearly
separated in \figref{pccd}(\emph{a}), have blended to form a continuous
distribution in \figref{pccd}(\emph{b}).  The main difference between Groups~I
and II, which both are characterized by the presence of OB-stars, is the
fraction of light contributed by older stellar populations.  Group~III and
IV pixels, defined as having 1.15\,\le\,\fuvu\lt\,1.65~mag \emph{after}
extinction correction, are affected differently by the extinction correction
(as a comparison of Figures~\ref{pccd}(\emph{a}) and \ref{pccd}(\emph{b})
 shows). 
Like Group~I and II pixels, Group~IV pixels are found scattered out to much
redder \bmir\ colors \emph{before} extinction correction.  Most of the
Group~III pixels, on the other hand, can be located in the same color--color
space in both panels of \figref{pccd}.  The same difference is seen between
pixel groups V and VI.  This indicates that Group~III and V pixels have
almost no measurable dust extinction, while Group~IV and VI pixels are
significantly obscured by dust.  As a side note, the reduction of the
scatter going from the observed to the extinction-corrected version of the
\bmir\ versus \fuvu\ pCCD is only possible if the extinction correction
applied to each individual pixel was appropriate, at least to first order. 

The pixel-averaged dust extinction estimated by our method in NGC\,0959 
(\figref{avmap}) indeed varies from \av\,=\,0.0 to \av\,$\simeq
$\,0.8\,\magarcsec.  In Paper~I, we also measured the average visual dust
extinction across the entire galaxy, $\overline{A_V}$\,=\,0.064\,$
^{+0.086}_{-0.049}$ \magarcsec, and the azimuthally averaged radial
extinction profile, $A_V{\rm (R)}$, which has a central value of $A_V{\rm
(R=0)}$\,$\simeq$\,0.25 \magarcsec, but quickly drops to below \av\,=\,0.1
\magarcsec\ beyond a radius of $\sim$\,0.2\,R$_{25}$, where R$_{25}$ denotes
the major axis radius at the $\mu_B$\,=\,25.0 \magarcsec\ isophote
\citep[RC3;][]{devaucouleurs91}.  These average \av\ values are much smaller
than the peak pixel-based \av\ estimate.  Also, while the $\overline{A_V}$
and radial profile do not become zero at any radius, with our method (see
Paper~I) about $\sim$\,55\% of the analyzed pixels in NGC\,0959 have
\av\,=\,0.0 \magarcsec\ to within the photometric uncertainties. 

Before we proceed to the next section, let us consider what might cause the
difference in dust extinction between Groups~I and III on the blue sequence. 
Dust and gas are easily removed by starburst heating
\citep[e.g.,][]{mihos94,mihos96} and stellar winds \citep[e.g,][]{murray05},
and both mechanisms become stronger as the size of an OB association becomes
larger.  Therefore, if the dominant stellar populations in Group~III pixels
are indeed evolved Group~I populations wherein OB-stars have died off, we
would \emph{expect} them to suffer less extinction, unless some mechanism
for rapid reformation and pervasive distribution of dust were at play. 
Figures~\ref{pccd}(\emph{a}) and \ref{pccd}(\emph{b}) show that, 
once dust-free,
these populations do \emph{not} become significantly dustier evolving from
Group~III to Group V, apparently ruling out such rapid reformation.  Since
we do not observe a similar strong drop in extinction going from Group~II to
Group~IV, we conclude that most of the extinction in those pixels is not
physically associated with the OB-associations themselves, although they
could still represent star formation ``blisters'' on the far side of large
molecular cloud complexes.

\subsection{Is NGC\,0959 Unique? Application to Other Galaxies}

To determine whether our pixel-based extinction correction can be used as a
general method, or whether NGC\,0959 is a special case, we applied the same
method to NGC\,7320 (SA(s)d at $D$\,=\,14.0\,$\pm$\,1.0\,Mpc, sampling
102\,$\times$\,102\,pc$^2$ pixel$^{-1}$) and UGC\,10445 (SBc at
$D$\,=\,20.0\,$\pm$\,1.4\,Mpc, 146\,$\times $\,146\,pc$^2$ pixel$^{-1}$),
which, although more distant, are selected from our larger sample of 45
galaxies with FUV through mid-IR imagery as relatively close analogs.  A
more detailed analysis of the distribution of extinction and the intrinsic
stellar populations of these galaxies will be performed in Tamura et al.\
(2010; Paper~III, in preparation).  Here we simply compare their \bmir\ 
versus \fuvu\ pCCDs \emph{before} and \emph{after} the application of 
our pixel-based extinction correction, to help validate the 
current results for NGC\,0959.

\begin{figure*} 
  \centerline{
  \includegraphics[width=0.49\textwidth,angle=0,clip=]{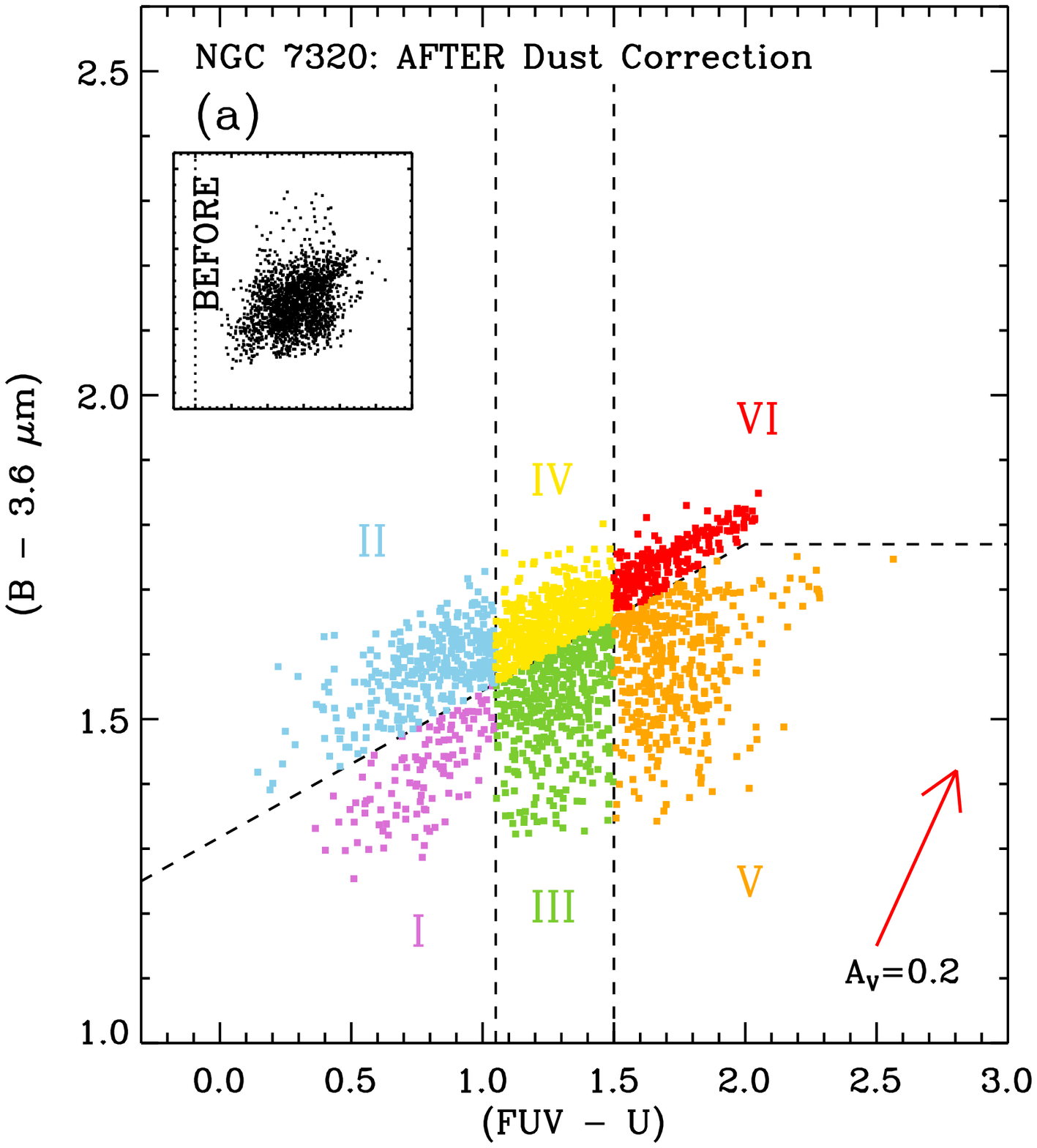} \
  \includegraphics[width=0.49\textwidth,angle=0,clip=]{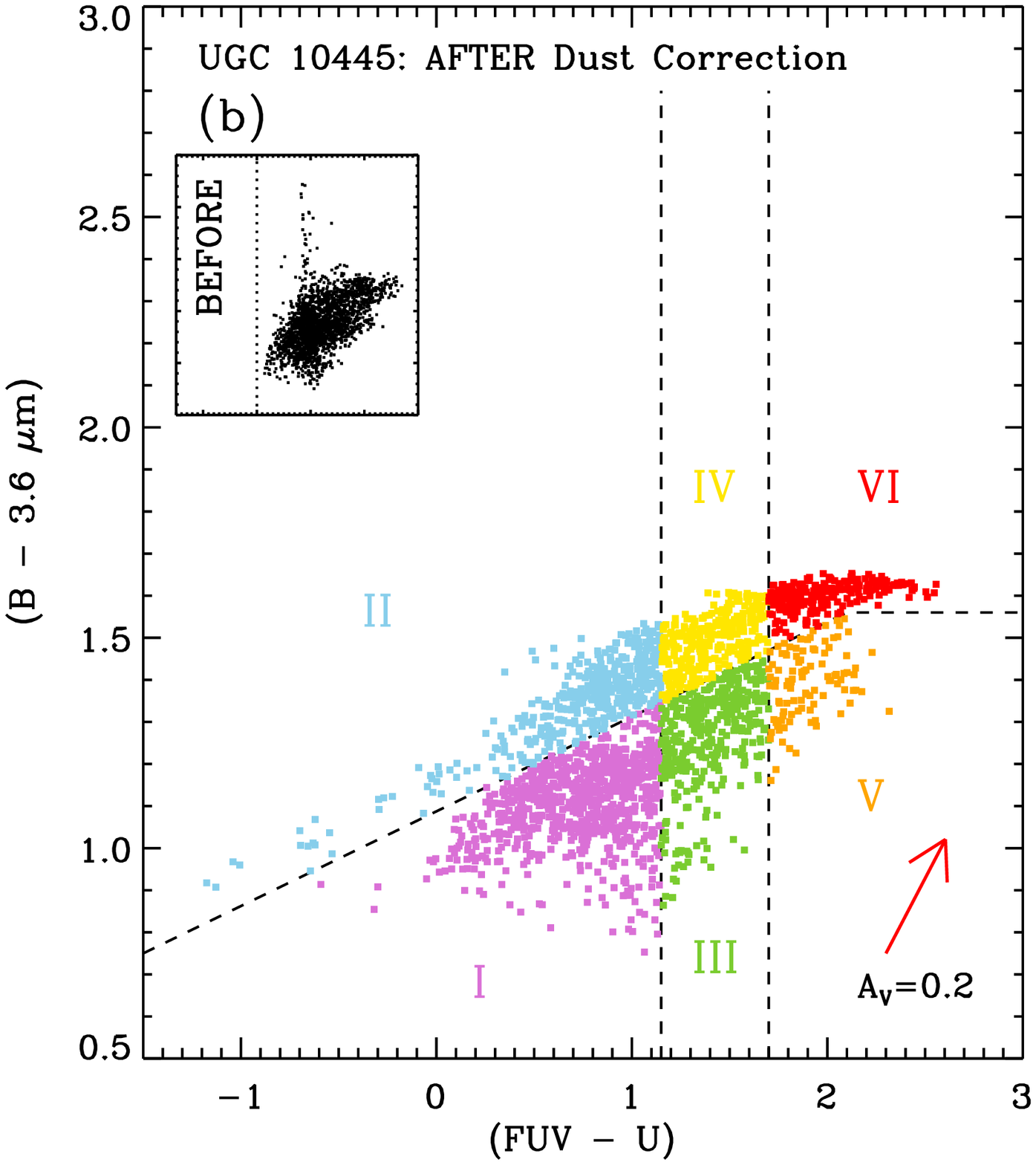}
  }
  \caption{\footnotesize
Extinction corrected pCCDs for NGC\,7320 (SA(s)d,
$D$\,=\,14.0\,$\pm$\,1.0\,Mpc) and UGC\,10445 (SBc,
$D$\,=\,20.0\,$\pm$\,1.4\,Mpc), two more distant analogs of NGC\,0959
($D$\,=\,9.9\,$\pm$\,0.7\,Mpc).  Since the area subtended by a single pixel
is (much) larger in these galaxies, the effects of light blending are more
severe, resulting in smaller and fuzzier separations of red and blue
sequences than for NGC\,0959.  Of course, their star formation histories may
also differ.  Yet, groupings of pixels that are qualitatively similar to
those in NGC\,0959 are recognizable in both pCCDs.  The insets, plotted over
the same color ranges as the main panels for each galaxy, show the
distribution of pixels before dust correction.  The vertical dotted lines in
the inset panels indicate the locations of \fuvu\,=\,0.0 mag, for comparison
with the main panels.
  }\label{ugcs}
\end{figure*}

\figref{ugcs} shows the distribution of the pixels of NGC\,7320 and
UGC\,10445 that meet our \SN\ criteria in the same color--color space as
\figref{pccd}.  The main panels show the distributions \emph{after}, the
insets show the distributions \emph{before} extinction correction.  For each
galaxy, both the main panel and inset show the \emph{same} ranges in color. 
The vertical dotted lines in the insets represent \fuvu\,=\,0.0\,mag. 
Reddening vectors corresponding to \av\,=\,0.2 \magarcsec\ are drawn in the
lower right corners of the main panels.  Since the area subtended by a
single pixel is larger than was the case for NGC\,0959, the blending of light
from distinct stellar populations within a single pixel becomes more
significant, resulting in less clear separation of blue and red sequences. 
Yet, different groupings of pixels, analogous to those defined in
\figref{pccd}(\emph{a}) for NGC\,0959, are still recognizable in 
\figref{ugcs} for both galaxies.

\subsection{A Pixel Coordinate Map of Stellar Populations within NGC\,0959}

\begin{figure} 
  \centerline{
  \includegraphics[width=0.49\textwidth,angle=0]{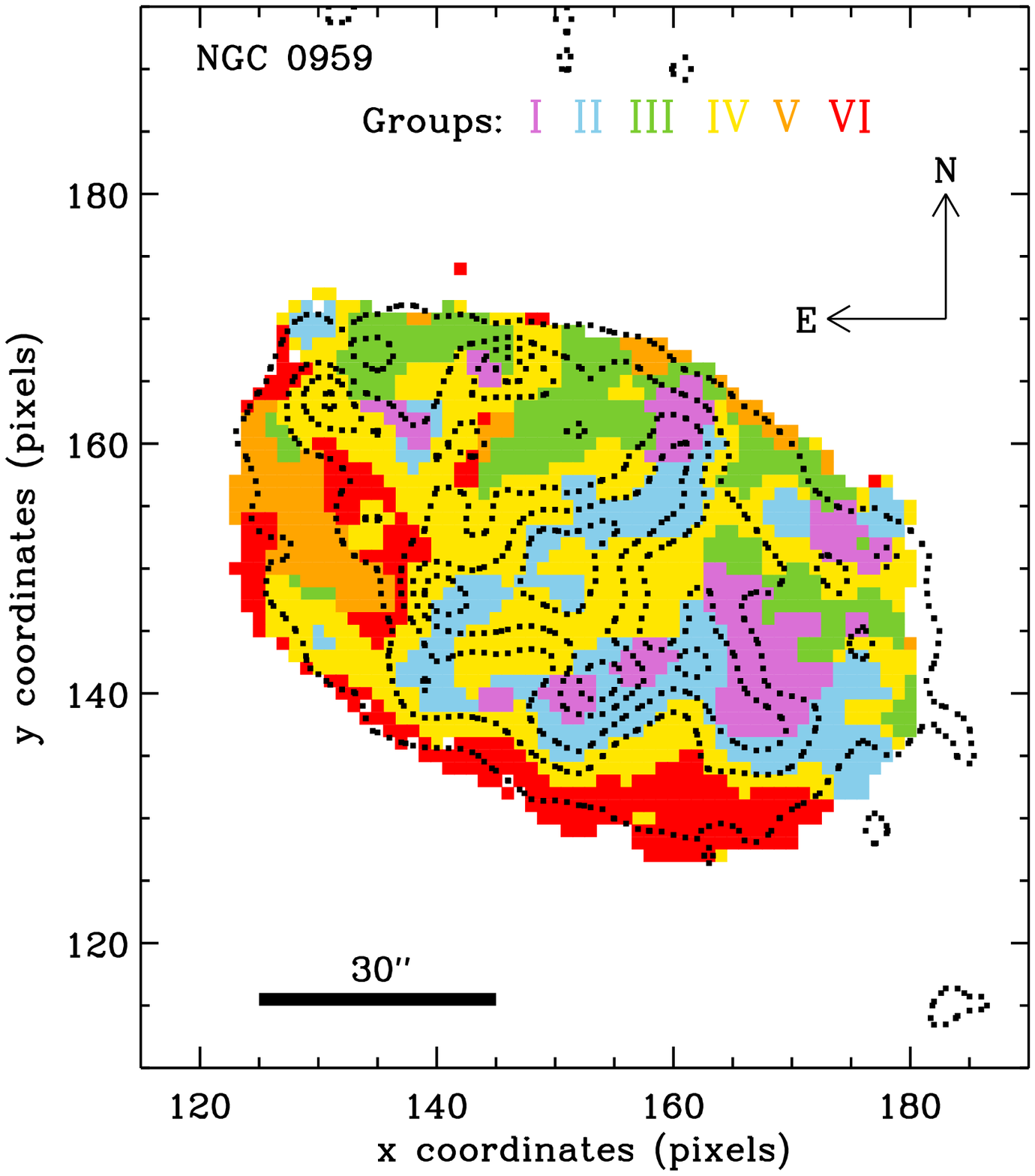}
  }
  \caption{\footnotesize
Pixel coordinate map of NGC\,0959, showing the spatial distribution of
pixels belonging to the different pixel groups selected in
\figref{pccd}(\emph{a}), i.e., \emph{after} application of our
pixel-based extinction correction.  Each square represents a single
1\farcs5$\times$1\farcs5 pixel.  The black dotted contours trace the
\emph{Spitzer}/IRAC 8.0\,\mum\ emission.  Group~I (purple) and
Group~II (blue) pixels appear to define some large-scale structure in
the galactic disk, highlighted by OB-associations.  Especially,
Group~II pixels reveal a \emph{previously unrecognized} bar-like
structure, running from NW to SE across the galaxy center.
  }\label{pixmap}
\end{figure}

Having shown that application of our pixel-based extinction correction
reveals significant groupings of pixels---dominated by different types or
mixtures of stellar populations---in the \bmir\ versus \fuvu\ pCCD, our 
next question becomes whether the spatial distribution of pixels belonging 
to these pixel groups will reveal meaningful large- and small-scale physical
structures within NGC\,0959.  To address this question, and to study how 
each pixel group relates to the visual and physical properties of stellar
populations, we plot the pixel groups defined in \figref{pccd}(\emph{a}) 
onto a two-dimensional pixel coordinate map (\figref{pixmap}).  Each 
square in \figref{pixmap}
represents a 1\farcs5\,$\times$\,1\farcs5 pixel, color-coded according to
the pixel group it is a member of.  Black dotted contours again trace the
\emph{Spitzer}/IRAC 8.0\,\mum\ PAH emission.  We emphasize that each pixel
group was defined \emph{without} using any pixel coordinate information. 
In other words, this is the \emph{first time} that the spatial distribution
of pixels belonging to each of the selected pixel groups is revealed. 

First and foremost, \figref{pixmap} demonstrates that the different pixel
groups are \emph{not} distributed randomly across the face of NGC\,0959's
galactic disk, but cluster in well-defined regions.  No systematics in the
data processing and process of defining the pixel groups would be expected
to result in spatial artifacts larger than 3$\times$3 pixels.  Since most
regions cover much larger contiguous areas, we conclude that also the
smallest spatial groupings must be genuine.  Since defining the different
pixel groups was possible only in the extinction corrected \bmir\ versus
\fuvu\ pCCD, a two-dimensional extinction correction is thus crucial to
reveal regions with systematically different star-formation histories. 

Group~I pixels (purple) are found in and around some of the bluest regions
in the extinction-corrected color composite of FUV, $V$, and 3.6\,\mum\
images (\figref{colimg}(\emph{b})), most of which are recognizable as such in
the uncorrected, higher resolution, ground-based \emph{UVR} color composite
(\figref{vattimg}) as well.  Since the stellar populations dominating these
pixels are much brighter than those in other regions (\figref{vattimg}),
these regions must indeed be vigorously star-forming. 

Group~II pixels (blue) are generally distributed in regions adjacent to the
Group~I pixels.  Each of the regions (b) and (c) of \secref{color_comp} 
are found to
belong to this pixel group, as well.  In particular, much of a more or less
linear structure that runs from the NW to SE of the galaxy through its
center, with high 8.0\,\mum\ surface brightness (PAH emission) and signs of
higher than average obscuration (Figures~\ref{avmap} and \ref{vattimg}),
consists of pixels that belong to Group~II.  
These regions have the blue
\fuvu\ colors characteristic of OB associations, but are redder in \bmir,
because a larger fraction of the light is contributed by older stellar
populations.  These observations are consistent with the presence of a
stellar bar.  Although NGC\,0959 has been studied before
\citep[e.g.,][]{esipov91, taylor05}, to our knowledge this bar has not
previously been reported.  The dust-corrected color composite in
\figref{colimg}(\emph{b}) also suggests that NGC\,0959 has a non-negligible
bulge or central condensation, that is partially obscured by dust in
Figures~\ref{colimg}(\emph{a}) and \ref{vattimg}.  We suggest, therefore, 
that its morphological classification be changed from Sdm
\citep[RC3;][]{devaucouleurs91} or Sc/Irr \citep[UGC;][]{nilson73} to SBcd. 

An interesting contrast in average stellar population age is revealed by
comparing the spatial distributions of Group~III (green), Group~IV (yellow)
and Group~V and VI (orange and red) pixels.  Whereas Group~III pixels are
mostly found in the northwestern half of the galaxy, Group~V and VI pixels
are distributed predominantly toward the eastern and southeastern periphery
of the galaxy (although a smaller region of Group~V pixels appears along the
northern rim).  The regions occupied by Group~III pixels appear neither
particularly blue (actively star-forming) nor red (quiescent) in
\figref{colimg}(\emph{b}), consistent with the idea that intermediate age 
(a few 100 Myr) stellar populations are the dominant contributors to the 
flux in these pixels.  The extinction map (\figref{avmap}) shows that 
these pixels suffer no, or at most minimal extinction by dust.  The Group~V
regions correspond to regions that have a smooth appearance without abrupt
changes in surface brightness in Figures~\ref{colimg} and \ref{vattimg}, 
and lacking any signatures of dust (Figures~\ref{avmap} and \ref{vattimg}).  
Their red \fuvu\ colors and neutral \bmir\ colors indicate that these must 
be regions with intermediate age ($\sim$1--2 Gyr), largely unattenuated 
stellar populations in the outskirts of the galaxy.  The Group VI pixels 
(red) in the southern half of the galaxy disk suffer some extinction 
(\figref{avmap}), but the low surface brightness in these regions does not
allow one to easily discern dust features in Figures~\ref{colimg} and
\ref{vattimg}.  Even though these regions suffer some extinction,
\figref{pccd}(\emph{a}) demonstrates that these pixels are red mostly 
because the light is dominated by older (likely older than a few Gyr) 
stellar populations.  The Group~IV pixels are distributed throughout the 
galaxy between the other populations, but perhaps mostly surrounding the 
Group~II pixels.  Like Group~III, their \fuvu\ colors indicate the presence 
of relatively young stars and absence of OB-stars, but the redder \bmir\ 
colors of Group~IV pixels show that a larger fraction of the flux is 
contributed by older populations.  The extinction map (\figref{avmap}) 
shows that these pixels tend to suffer
somewhat less attenuation than Group~I and II pixels.  The overall
impression is that of a large-scale past star-formation episode that started
in the south or southeast and propagated toward the northwest, and that is
unrelated to the ongoing massive star formation traced by Group~I and II
pixels, which presently is concentrated more toward the western half of
NGC\,0959.  

The apparent ridge of somewhat higher extinction along the southeastern edge
(as defined by our \SN\ge3 criterion) of NGC\,0959, so striking in the 
extinction map (\figref{avmap}), might hint at the presence of 
either a warp in the galactic disk or an outer spiral arm delineated by a 
dust lane.  In either case, no significant star-formation must have been 
associated with that portion of the galactic disk for the past few Gyr. 

Concerning the SE-to-NW stellar population gradient of NGC\,0959 and its
SE extinction ridge, it is worth noting that NGC\,0959 is a member of the
NGC\,1023 galaxy group, which contains 12 other major galaxies 
\citep{tully80}.  The center of that group is located $\sim$1.6$^{\circ}$
($\sim$240\,kpc at $D$\,$\simeq$\,10\,Mpc) south of NGC\,0959.  The
group environment and/or gravitational interaction with NGC\,0949---the
group member nearest in projection on the sky---may well have played a
role in producing the extinction ridge and the stellar population
gradient in the disk of NGC\,0959.  We believe this certainly deserves 
further study.

\section{Summary and Conclusions}

We have presented the results of a study of NGC\,0959 using color composite
images, a pCCD, and a pixel coordinate map, and demonstrated the importance
and potential of a pixel-based extinction correction.  Our study combined
ground- and space-based surface photometry, ranging in wavelength from the
far-UV (\emph{GALEX}) through mid-IR (\emph{Spitzer}/IRAC).  Among the
possible combinations of color-magnitude and color-color diagrams examined,
we found that the \bmir\ versus \fuvu\ pCCD proved the most powerful
diagnostic of differences in the stellar populations contributing to the
flux in a pixel.  After applying the pixel-based, two-dimensional extinction
correction described in Paper~I, we defined six different pixel groups in
this diagnostic diagram, ranging from pixels for which both \fuvu\ and
\bmir\ colors indicate that their fluxes are dominated by very young,
massive stellar populations, to pixels that appear to sample only light from
evolved stellar populations.  We demonstrated that it was not possible to
meaningfully define such pixel groups \emph{before} 
extinction correction.  We then
showed that pixels that belong to a given pixel group form well-defined,
contiguous regions in a pixel coordinate map, revealing systematic spatial
variations in the dominant stellar populations that would not be readily
discernible without a two-dimensional correction for extinction.  We were
able to report the presence of a previously unrecognized stellar bar, for
example. 

Our pixel-based two-dimensional method to correct for extinction, based on
only 3.6\,\mum\ and $V$-band images (adding $U$ improves fidelity), has the
advantage that it is applicable to any galaxy that is significantly resolved
at rest-frame $\sim$3.6\,\mum\ and that it is not restricted to a few
individual sightlines within a galaxy, nor to the very nearest galaxies. 
Although seemingly simple and crude, we have demonstrated that application
of this method allowed us to uncover relatively detailed spatial information
on the nature of stellar populations and on the star formation history
within NGC\,0959, despite significant spatial variations in the attenuation
by intervening dust. 

While the spatial resolution in the present study was limited by the
\emph{GALEX} NUV PSF of 5\farcs3 (FWHM) to linear scales of $\sim$250\,pc,
in the near future, combination of \emph{HST}/WFC3 and \emph{JWST} will 
provide much higher resolution (FWHM\,$\simeq$\,0\farcs06--0\farcs11) at
rest-frame UV through mid-IR wavelengths.  These instruments will thus allow
us to study the distribution of dust extinction and the underlying stellar
populations at high fidelity and at much finer spatial resolution within
nearby galaxies, as well as at similar scales of a few hundred pc within
more distant galaxies.  Since galaxies at $z$\,$\gtrsim$\,1 are important
building blocks of the galaxies that we see today at $z$\,$\simeq$\,0, 
studying these higher redshift objects with proper two-dimensional 
extinction estimates will be of significant importance to reveal the 
nature of galaxy assembly and evolution.

\vspace{0.5cm}
{\small 

This work is funded by NASA/ADP grant NNX07AH50G and NASA/JWST grant
NAG\,5-12460.  R.A.W.\ was supported in part by NASA/JWST grant
NAG\,5-12460.  We thank Violet Mager for providing the data observed at the
Vatican Advanced Technology Telescope (VATT): the Alice P.\ Lennon Telescope
and the Thomas J.\ Bannan Astrophysics Facility.  We also thank Daniela
Calzetti for useful discussions that helped improve our paper.  This study
has made use of the NASA/IPAC Extragalactic Database (NED), which is
operated by the Jet Propulsion Laboratory (JPL), California Institute of
Technology, under contract with the National Aeronautics and Space
Administration (NASA).  This study has also made use of NASA's Astrophysics
Data System (ADS).  We thank the anonymous referee for constructive comments
that helped improve this paper. 

}

\renewcommand{\baselinestretch}{0.90}

\clearpage\renewcommand{\baselinestretch}{1.111111}


{\footnotesize
\begin{thebibliography}{}
\bibitem[Anders \& Fritze-von Alvensleben, 2003]{anders03} Anders, P., \& 
  Fritze-von Alvensleben, U.\ 2003, A\&A, 401, 1063
\bibitem[Boissier \etal, 2004]{boissier04} Boissier, S., Boselli, A., 
  Buat, V., Donas, J., Milliard, B.\ 2004, A\&A, 424, 465
\bibitem[Boissier \etal, 2005]{boissier05} Boissier, S., \etal\ 2005, ApJ, 
  619, L83
\bibitem[Bruzual \& Charlot, 2003]{bruzual03} Bruzual, G., \& Charlot, S.\ 
  2003, MNRAS, 344, 1000
\bibitem[Buat \& Xu, 1996]{buat96} Buat, V., \& Xu, C.\ 1996 A\&A, 306, 61
\bibitem[Byun, 1992]{byun92} Byun, Y.-I.\ 1992, Ph.D.\ thesis, Australian
  National University
\bibitem[Calzetti \etal, 1994]{calzetti94} Calzetti, D., Kinney, A., L.,
  \& Storchi-Bergmann, T.\ 1994, ApJ, 429, 582
\bibitem[Calzetti \etal, 2000]{calzetti00} Calzetti, D., Armus, L., 
  Bohlin, R.\ C., Kinney, A.\ L., Koornneef, J., \& Storchi-Bergmann, T.\ 
   2000, ApJ, 533, 682
\bibitem[Calzetti \etal, 2005]{calzetti05} Calzetti, D., \etal\ 2005, 
  ApJ, 633, 871
\bibitem[Calzetti \etal, 2007]{calzetti07} Calzetti, D., \etal\ 2007, 
  ApJ, 666, 870
\bibitem[Caplan \& Deharveng, 1985]{caplan85} Caplan, J., \& Deharveng, L.\
  1985, A\&AS, 62, 63
\bibitem[Clayton \etal, 2000]{clayton00} Clayton, G.\ C., 
  Gordon, K.\ D., \& Wolfe, M.\ J.\ 2000, ApJS, 129, 147
\bibitem[Deo \etal, 2006]{deo06} Deo, R., P., Crenshaw, D., M., \& 
  Kraemer, S., B.\ 2006, AJ, 132, 321
\bibitem[de Vaucouleurs \etal, 1991]{devaucouleurs91} de Vaucouleurs, G.,
  de Vaucouleurs, A., Corwin, H.\ G., Buta, R.\ J., Paturel, G., \& 
  Fouque, P.\ 1991, Third Reference Catalogue of Bright Galaxies 
  (Springer, New York) (RC3)
\bibitem[Driver \etal, 2008]{driver08} Driver, S.\ P., Popescu, C.\ C., 
  Tuffs, R.\ J., Graham, A.\ W., Liske, J., \& Baldry, I.\ 2008, ApJ,
  678, 101
\bibitem[Elmegreen, 1980]{elmegreen80} Elmegreen, D.\ M.\ 1980, ApJS, 43, 37
\bibitem[Esipov \etal, 1991]{esipov91} Esipov, V.\ F., Kyazumov, G.\ A., \& 
  Dzhafarov, A.\ R.\ 1991, SvA, 35, 452
\bibitem[Fazio \etal, 2004]{fazio04} Fazio, G.\ G., \etal, 2004, ApJS, 
  154, 10
\bibitem[Gil de Paz \etal, 2007]{gildepaz07} Gil de Paz. A., \etal\ 2007, 
  ApJS, 173, 185
\bibitem[Gordon \& Clayton, 1998]{gordon98} Gordon, K.\ D., \& 
  Clayton, G.\ C.\ 1998, ApJ, 500, 816
\bibitem[Gordon \etal, 2003]{gordon03} Gordon, K.\ D., Clayton, G.\ C.,
  Misselt, K.\ A., Landolt, A.\ U., \& Wolff, M.\ J.\ 2003, ApJ, 594, 279
\bibitem[Kaviraj \etal, 2007]{kaviraj07} Kaviraj, S., Rey, S.-C., Rich, R.\ 
  M., Yoon, S.-J., \& Yi, S\ K.\ 2007. MNRAS, 381, L74
\bibitem[Kennicutt \etal, 2009]{kennicutt09} Kennicutt, R.\ C., \etal\
  2009, ApJ, 703,1672
\bibitem[Kong \etal, 2004]{kong04} Kong, X., Charlot, S., Brinchmann, C., 
  \& Fall, S.\ M.\ 2004, MNRAS, 349, 769
\bibitem[Kotulla \etal, 2009]{kotulla09} Kotulla, R., Fritze, U., 
  Weibacher, P., \& Anders, P.\ 2009, MNRAS, 396, 462
\bibitem[Maraston, 2005]{maraston05} Maraston, C.\ 2005, MNRAS, 362, 799
\bibitem[Martin \etal, 2005]{martin05} Martin, D.\ C., \etal\ 2005, ApJ, 619, 
  L1
\bibitem[Mathis \etal, 1977]{mathis77} Mathis, J.\ S., Rumpl, W., \& 
  Nordsieck, K.\ H.\ 1977, ApJ, 217, 425
\bibitem[Mihos \& Hernquist, 1996]{mihos96} Mihos, C.\ J., \& Hernquist, L.\
  1996, ApJ, 464, 641
\bibitem[Mihos \& Hernquist, 1994]{mihos94} Mihos, C.\ J., \& Hernquist, L.\
  1994, ApJ, 427, 112
\bibitem[Misselt \etal, 1999]{misselt99} Misselt, K.\ A., Clayton, 
  G.\ C., \& Gordon, K.\ D.\ 1999, ApJ, 515, 128
\bibitem[Morrissey \etal, 2007]{morrissey07} Morrissey, P., \etal\ 2007,
  ApJS, 173, 682
\bibitem[Mould \etal, 2000]{mould00} Mould, J.\ R., \etal\ 2000, ApJ, 
  529, 786 
\bibitem[Murray \etal, 2005]{murray05} Murray, N., Quataert, E., \& 
  Thompson, T.\ A.\ 2005, ApJ, 618, 569
\bibitem[Nilson, 1973]{nilson73} Nilson, P.\ 1973, Uppsala General Catalogue
  of galaxies, Uppsala Obs.\ Ann., vol.\ 6 (UGC)
\bibitem[Oke, 1974]{oke74} Oke, J.\ B.\ 1974, ApJS, 27, 21
\bibitem[Oke \& Gunn, 1983]{oke83} Oke, J.\ B., \& Gunn, J.\ E., 1983, ApJ,
  266, 713
\bibitem[Petersen \& Gammelgaard, 1997]{petersen97} Petersen, L., \& 
  Gammelgaard, P.\ 1997, A\&A, 323, 697
\bibitem[Price \etal, 2002]{price02} Price, S.\ D., Carey, S.\ J., \& 
  Egan, M.\ P.\ 2002, AdSpR, 30, 2027
\bibitem[Regan, 2000]{regan00} Regan, M.\ W., 2000, ApJ, 541, 142
\bibitem[Rieke \etal, 2004]{rieke04} Rieke, G.\ H., \etal\ 2004, ApJS, 
  154, 25
\bibitem[Roussel \etal, 2005]{roussel05} Roussel, H., Gil de Paz, A., 
  Seibert, M., Helou, G., Madore, B.\ F., \& Martin, C.\ 2005, ApJ, 
  632, 227
\bibitem[Rudy, 1984]{rudy84} Rudy, R.\ J.\ 1984, ApJ, 284, 33
\bibitem[Scoville \etal, 2001]{scoville01} Scoville, N.\ Z., Polletta, 
  M., Ewald, S., Stolovy, S.\ R., Thompson, R., \& Rieke, M.\  2001, AJ, 
  122, 3017
\bibitem[Skrutskie \etal, 2006]{skrutskie06} Skrutskie, M.\ F., Cutri, R.\
  M., Stiening, R., \etal\ 2006, AJ, 131, 1163
\bibitem[Tamura \etal, 2009]{tamura09a} Tamura, K., Jansen, R.\ A., 
  Windhorst, R.\ A.\ 2009, AJ, 138, 1634  (Paper~I)
\bibitem[Taylor \etal, 2005]{taylor05} Taylor, V.\ A., Jansen, R.\ A., 
  Windhorst, R.\ A., Odewahn, S.\ C., \& Hibbard, J., E.\ 2005, ApJ, 
  630, 784
\bibitem[Trumpler, 1930]{trumpler30} Trumpler, R.\ J.\ 1930, PASP, 42, 214
\bibitem[Tully, 1980]{tully80} Tully, R.\ B.\ 1980, ApJ, 237, 390
\bibitem[van Houten, 1961]{vanhouten61} van Houten, C.\ J.\ 1961, Bull.\
  Astron.\ Inst.\ Neth., 16, 1
\bibitem[Valencic \etal, 2003]{valencic03} Valencic, L.\ A., 
  Clayton, G.\ C., Gordon, K.\ D., \& Smith, T.\ L.\ , 2003, ApJ, 598, 369
\bibitem[Viallefond \etal, 1982]{viallefond82} Viallefond, F., Goss, W.\ M., 
  \& Allen R.\ J.\ 1982, A\&A, 115, 373
\bibitem[Waller \etal, 1992]{waller92} Waller, W.\ H., Gurwell, M., \& 
  Tamura, M.\ 1992, AJ, 104, 63
\bibitem[Walterbos \& Kennicutt, 1988]{walterbos88} Walterbos, R.\ A.\ M.,
  \& Kennicutt, R.\ C.\ 1988, A\&A, 198, 61
\bibitem[Whittet \etal, 2001]{whittet01} Whittet, D.\ C.\ B., Gerakines,
  P.\ A., Hough, J.\ H., \& Shenoy, S.\ S.\ 2001, ApJ, 547, 872
\bibitem[Whittet \etal, 2004]{whittet04} Whittet, D.\ C.\ B., Shenoy, S.\ S.,
  Clayton, G.\ C., \& Gordon, K.\ D.\ 2004, ApJ, 602, 291
\bibitem[Willner \etal, 2004]{willner04} Willner, S.\ P., \etal\ 2004, 
  ApJS, 154, 222
\bibitem[Witt \& Gordon, 2000]{witt00} Witt, A.\ N., \& Gordon, K.\ D., 
  2000, ApJ, 528, 799 
\bibitem[Witt \& Gordon, 1996]{witt96} Witt, A.\ N., \& Gordon, K.\ D., 
  1996, ApJ, 463, 681
\bibitem[Witt \etal, 1992]{witt92} Witt, A.\ N., Thronson, H.\ A., 
  \& Capuano, J.\ M.\ 1992, ApJ, 393, 611
\bibitem[Worthey, 1994]{worthey94} Worthey, G.\ 1994, ApJS, 95, 107
\bibitem[Xu \& Helou, 1996]{xu96} Xu, C., \& Helou, G.\ 1996, ApJ, 456, 152
%
\end{thebibliography}
}
\end{document}